\documentclass{emulateapj}

\newcommand{\wlam}{$W_{\lambda}$}

\slugcomment{To appear in AJ.}

\shorttitle{Intraday Variability in PKS~2005--489}
\shortauthors{Rector \& Perlman}

\begin{document}


\title{A Search for Intraday Variability in the Blazar PKS~2005--489}


\author{Travis A. Rector}
\affil{National Radio Astronomy Observatory, P.O. Box O, Socorro, NM 87801}
\email{trector@nrao.edu}

\and

\author{Eric S. Perlman}
\affil{Joint Center for Astrophysics, Physics Department, University of Maryland, Baltimore County, 1000 Hilltop
Circle, Baltimore, MD  21250}




\begin{abstract}

Rapid, multi-color optical monitoring carried out for a multiwavelength campaign during 22-27 August 2000 is
presented for the bright and variable BL Lac object PKS~2005--489.  
Previous multiwavelength monitoring campaigns of PKS~2005--489 have found complex variability behavior both in flux density
and in broadband spectrum, with different modes and timescales from days to hours.  During this
campaign, PKS~2005--489 was observed to be in an unusual state, bright in the optical but faint in the X-ray, just prior
to an X-ray flare that began on 2 September 2000.  PKS~2005--489 was also  observed to have a flatter optical spectrum than prior epochs.  A
linear increase in optical flux density was observed during the first 5.2 days of the campaign.  However, intraday and
color variability were not detected.  The large discrepancy between the ratio of X-ray to optical flux
density observed during this campaign and that observed in prior epochs indicates a lack of long-term correlation between the two
wavebands; and it illustrates that simple particle injection events in the jet are insufficient to explain the long-term variability behavior of this object.
Twenty-three in-field standard stars are identified for future optical monitoring programs.

\end{abstract}




\section{Introduction}

Multiwavelength monitoring campaigns of BL Lacs find complex variability behavior, involving different modes and
likely reflecting a complex mix of jet structure and emission mechanisms.  Like other classes of blazars, the
emission from BL Lacs is dominated from radio to soft X-ray energies by synchrotron emission from an energetic
plasma that is moving relativistically towards the observer.  At higher energies the emission is likely due to
inverse-Compton scattering of low-energy seed photons, either synchrotron emission from the jet itself or from external
sources.  Intraday variability (IDV), which is typically defined as variability on timescales of
$\sim$1 day in the source restframe,  is a common characteristic of BL Lac objects \citep{hei96}, although in high-energy peaked BL Lacs (HBLs) IDV is less common \citep{hei98}.  From causality
arguments one can infer that the emission regions in BL Lacs are smaller than light-days modulo beaming effects, or
roughly the size of the Solar System.  From observed flux densities and variability timescales, brightness
temperatures greater than the ``Compton limit" of $T_b > 10^{12}$ K \citep{kel69} are implied, requiring either
relativistic beaming or extrinsic (propagation induced) sources of variability.  The observed variability of many
sources at multiple wavelengths indicates that at least some of the variability is intrinsic, implying extreme
photon densities within the source as well as relativistic beaming \citep[and references therein]{wag95}.    

PKS~2005--489 is a bright and highly variable BL Lac object.  It is a member of the ``1 Jansky" sample of BL
Lacs \citep{sti91}, which is defined by its bright radio flux density ($S_{\rm 5 GHz} \geq 1$ Jy), flat radio
spectrum ($\alpha_r \geq -0.5$; $S_{\nu} \propto \nu^{+\alpha}$) and lack of strong emission lines (\wlam\ $\leq
5$\AA).  Despite is presence in the radio-bright 1Jy sample, PKS~2005--489 is classified as an intermediate to
high-energy peaked BL Lac object because its broadband spectrum peaks in the UV.  
Despite the lack of detection while in a
non-flaring state during our campaign \citep{rec02}, its proximity ($z=0.071$) and spectral energy distribution make
PKS~2005--489 a promising TeV candidate
\citep{per99a,cos02}.

Its bright X-ray flux has allowed PKS~2005--489 to be monitored in the X-ray in detail on several occasions; and it
has exhibited complex variability behavior.  For example, in September 1998, there was a general increase in
X-ray flux density over a 3-day period, with smaller oscillations on timescales of several hours
\citep{sam00}.  In October--December 1998 there was a massive flare, which peaked at F(2--10~keV) =
$3.33\times10^{-10} {\rm ~erg ~cm^{-2} ~s^{-1}}$.  During the flare, no variability was detected on
intraday timescales, and there was very little shift in $\nu_{peak}$ at $\sim 10^{15.5}-10^{16}$ Hz 
\citep{per99b,tag01}.

Based upon the past behavior of PKS~2005--489, we designed a multiwavelength monitoring campaign with the
following schedule.  Short ($\sim 2-3$ orbit) observations were taken with the {\it Rossi} X-Ray Timing
Explorer (RXTE) satellite at intervals of $\sim 0.5-1$ day during 19--26 August and 3--4 September 2000.
Nearly continuous RXTE monitoring was completed from 27 August to 2 September 2000.  Continuous optical
monitoring was scheduled for 22--28 August at Cerro Tololo, Chile, Adelaide, Australia and Potchefstroom,
South Africa.  Unfortunately it was not possible to schedule the X-ray and optical monitoring campaigns to be coincident.
PKS~2005--489 was also observed in the submillimeter by the Swedish--ESO Submillimetre Telescope (SEST) and at TeV
energies with CANGAROO-II.  The overall results of the multiwavelength campaign are presented in \citet{rec02}.  Here we
present a search for IDV from the optical monitoring portion of the campaign.

\section{Observation and Reduction}

The optical observations for the multiwavelength campaign were designed to search for variability on 
timescales from minutes to days.  Optical monitoring was completed with the CTIO 0.9\,m telescope during 22--28
August.  Unfortunately, optical monitoring could not be completed in Adelaide and in Potchefstroom due to bad
weather, thereby leaving $\sim$16 hour holes per day in the optical coverage. In addition, the first two nights
at CTIO were hampered by bad weather.  

The monitoring was completed in three colors by using the CTIO B Tek (4201/1050), V Tek (5475/1000) and R Tek
(6425/1500) filters.  These filters are roughly equivalent to Johnson B, Johnson V and Cousins R respectively.
Short exposures were taken in immediate succession.  Exposure times varied from 30 to 150 seconds depending on
weather conditions.  Including the CCD read-out time, a cycle of $BVR$ observations was completed roughly every
five minutes during photometric conditions.  PKS~2005--489 was observable for roughly ten hours per night,
allowing over 100 datapoints per filter to be obtained on most nights.

The CCD frames were overscan, bias and flat-field corrected in the standard manner.  The images were shifted
between exposures to avoid any systematic errors in the pixel-to-pixel calibration.  Aperture photometry was
completed on PKS~2005--489 and in-field stars using the IRAF APPHOT package as described in
\citet{mas92}.  An aperture of 15 pixels (6\arcsec) in diameter was chosen.

\begin{figure}
\plotone{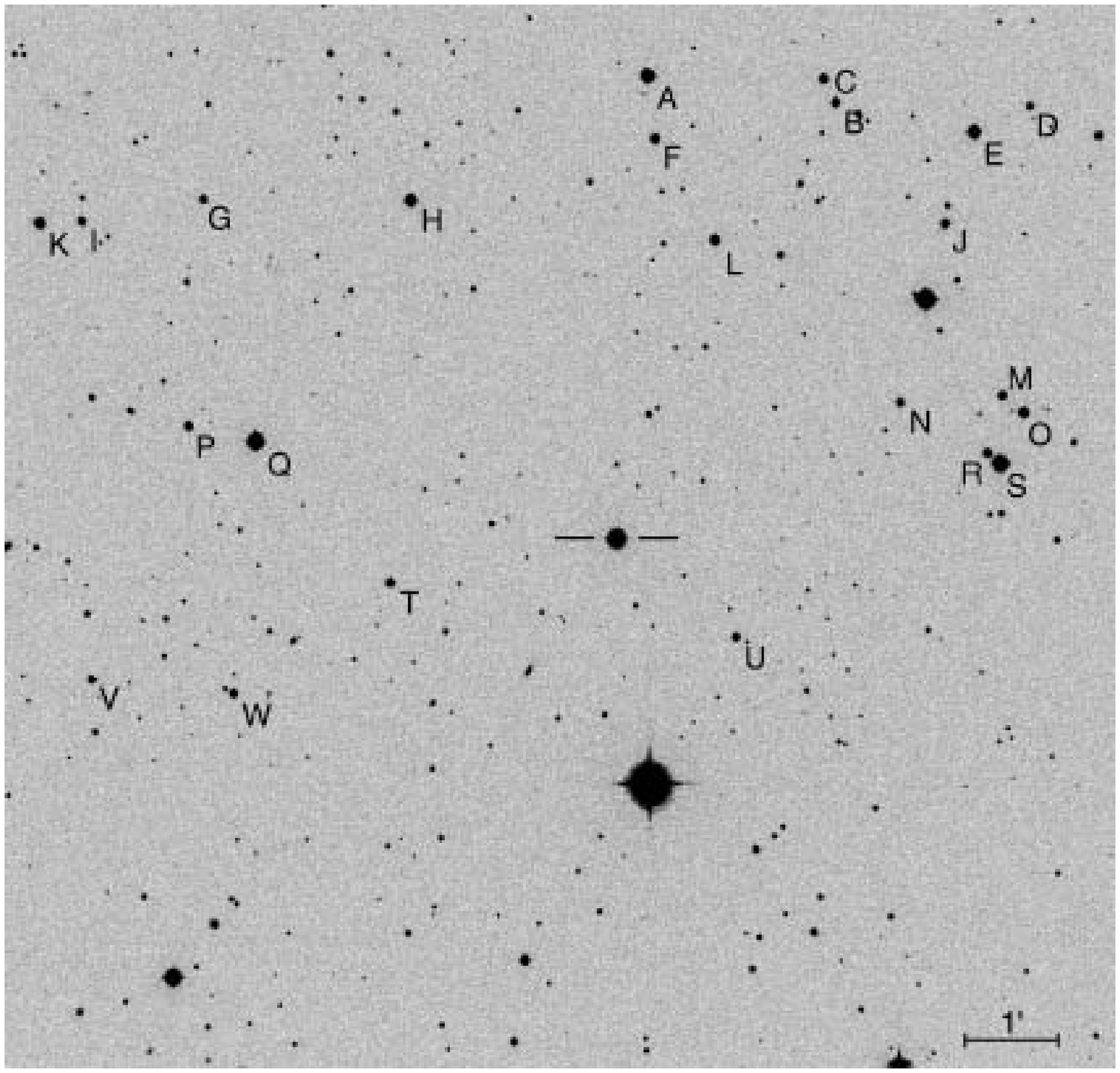}
\caption {The optical field of PKS~2005--489.  North is up and east to the left.  PKS~2005--489 is the marked
object to the right of the center of the image.  In-field standard stars are marked with the letters A through
W.}
\label{fig-1}
\end{figure}

The absolute photometric solutions were determined from multiple observations of Baldwin-Stone southern
hemisphere secondary spectrophotometric standards in \citet{lan92}.  The solutions were determined independently
for each photometric night and found to be consistent to $\la 0.05$ magnitudes for in-field standard stars.  
Twenty-three in-field standard stars were identified based upon their photometric stability during the seven
nights of the campaign.  These standards are shown in the optical field of PKS~2005--489 in Figure~\ref{fig-1} and 
a summary of their properties is given in Table~\ref{tbl-1}.  The columns are: [1] the label in
Figure~\ref{fig-1};  [2--3] the Right Ascension and Declination in J2000 coordinates.  The coordinates are accurate
to better than an arcsecond; [4]
the
$V$ magnitude; and [5--6] the \bv\ and \vr\ colors.  Magnitude values are limited by the consistency of the nightly
photometric solutions, and are therefore accurate to $\la 0.05$.  

The magnitude and colors of PKS~2005--489 are determined relative to the in-field standards, resulting in an
absolute photometric accuracy of $\sim 0.05$ magnitudes and an intraday relative photometric precision of $\la
0.015$ magnitudes.  The measured flux densities are corrected for Galactic reddening assuming $N_H =
5\times10^{20} {\rm ~cm^{-2}}$, which is consistent with {\it BeppoSAX} and {\it ROSAT} observations
\citep{pad01,sam95}.  Extinction coefficients of
$A_B = 0.351$,
$A_V = 0.265$ and
$A_R = 0.198$ are derived via the methodology in
\citet{car89}.  The flux density from the host galaxy, $m_r = 14.52\pm0.01$
\citep{urr00}, is also subtracted from PKS~2005--489 by assuming standard elliptical colors of \bv\ $=0.96$ and
\vr\ $=0.61$.

\section{Analysis}

Figure 2 of \citet{rec02} shows the optical variability of PKS~2005--489 over the entire campaign, during which
time the optical flux density increased linearly 13\% over 5.2 days ($\sim$500 ksec), starting 22 August,
and then remained constant during the night of 28 August.  The X-ray flux doubled over roughly the same time period
as the optical campaign \citep{rec02}.  During this campaign PKS~2005--489 was seen in an optically bright state, with a mean magnitude of $V=12.456\pm0.048$ and mean optical colors of $B-V=0.249\pm0.010$ and $V-R=0.230\pm0.006$.  This is $\sim$1
magnitude brighter than in quiescent epochs
\citep{wal86,hei96}.  There is no evidence of optical color variability during the campaign; however there is evidence for
long-term color variability. 
\citet{wal86} report 
$B-R = 0.84\pm0.06$ in September 1981 and 
$B-R = 0.88\pm0.06$ in August 1982, which is redder than the $B-R \approx 0.48$ observed during this campaign. 
The trend for the optical spectrum to flatten during bright phases has been observed in other BL Lacs as well,
e.g., S5 0716+714 \citep{wag96}.  The observed hardening of the spectrum may have been a precursor to the X-ray flare
which began on 2 September 2000, unfortunately four days after the optical campaign had concluded \citep{rec02}.

\begin{figure}
\plotone{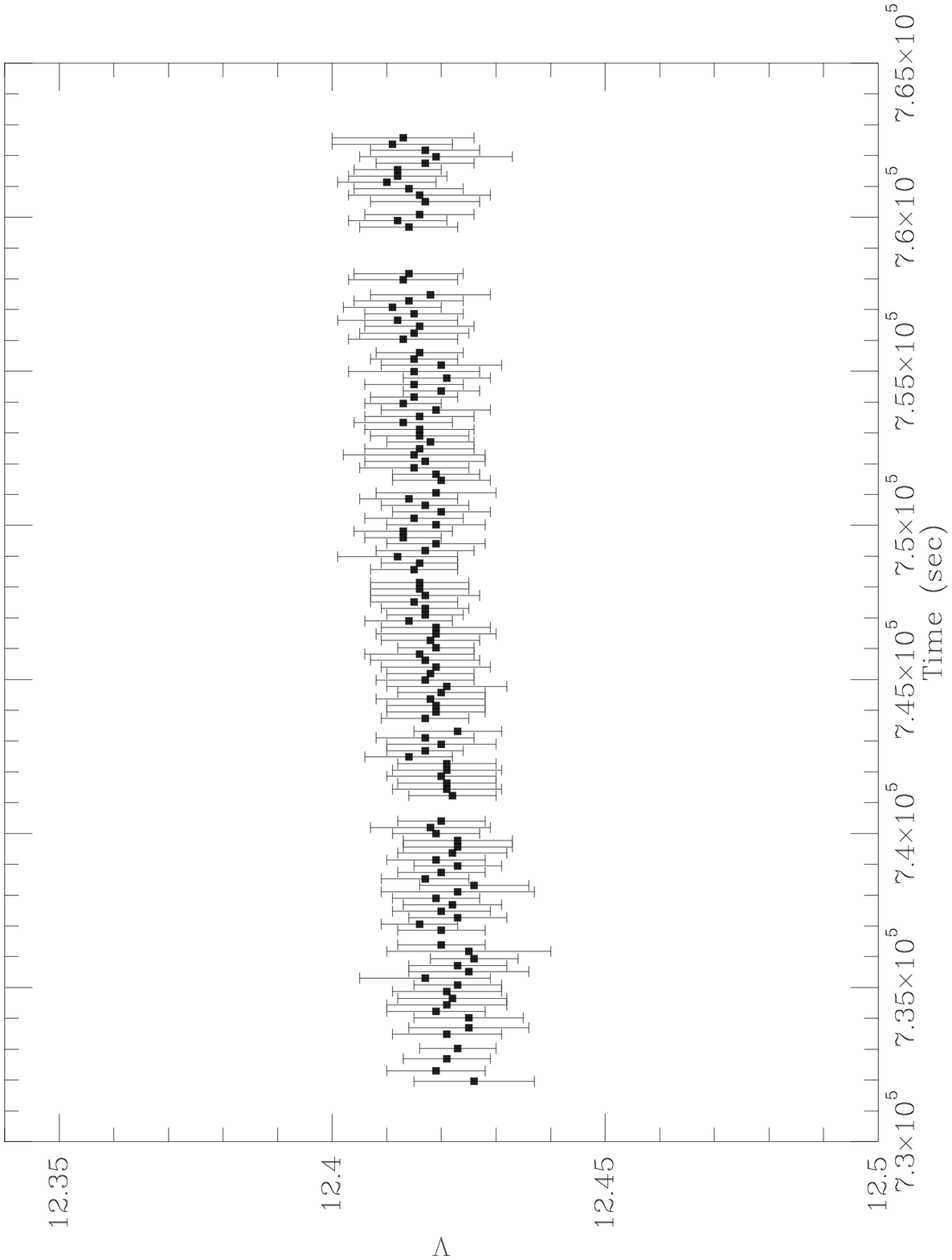}
\caption {The $V$ magnitude of PKS~2005--489 during the sixth night of the campaign. Time $T=0$ corresponds to
the beginning of the multiwavelength campaign on 19 August 2000 at 11:53:04 UTC.}
\label{fig-2}
\end{figure}

\begin{figure}
\plotone{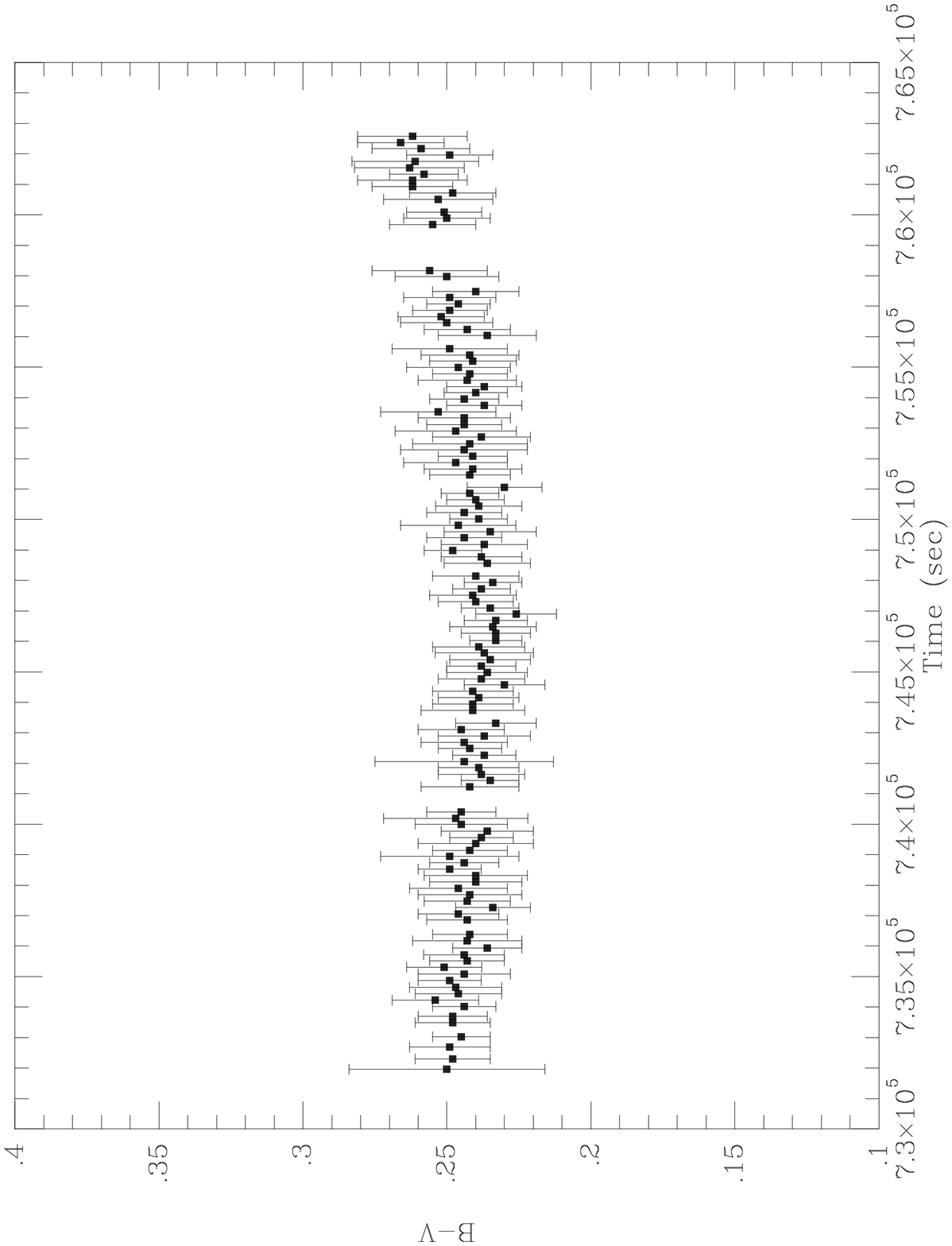}
\caption {The \bv\ color of PKS~2005--489 during the sixth night of the campaign. Time $T=0$ corresponds to
the beginning of the multiwavelength campaign on 19 August 2000 at 11:53:04 UTC.}
\label{fig-3}
\end{figure}

\begin{figure}
\plotone{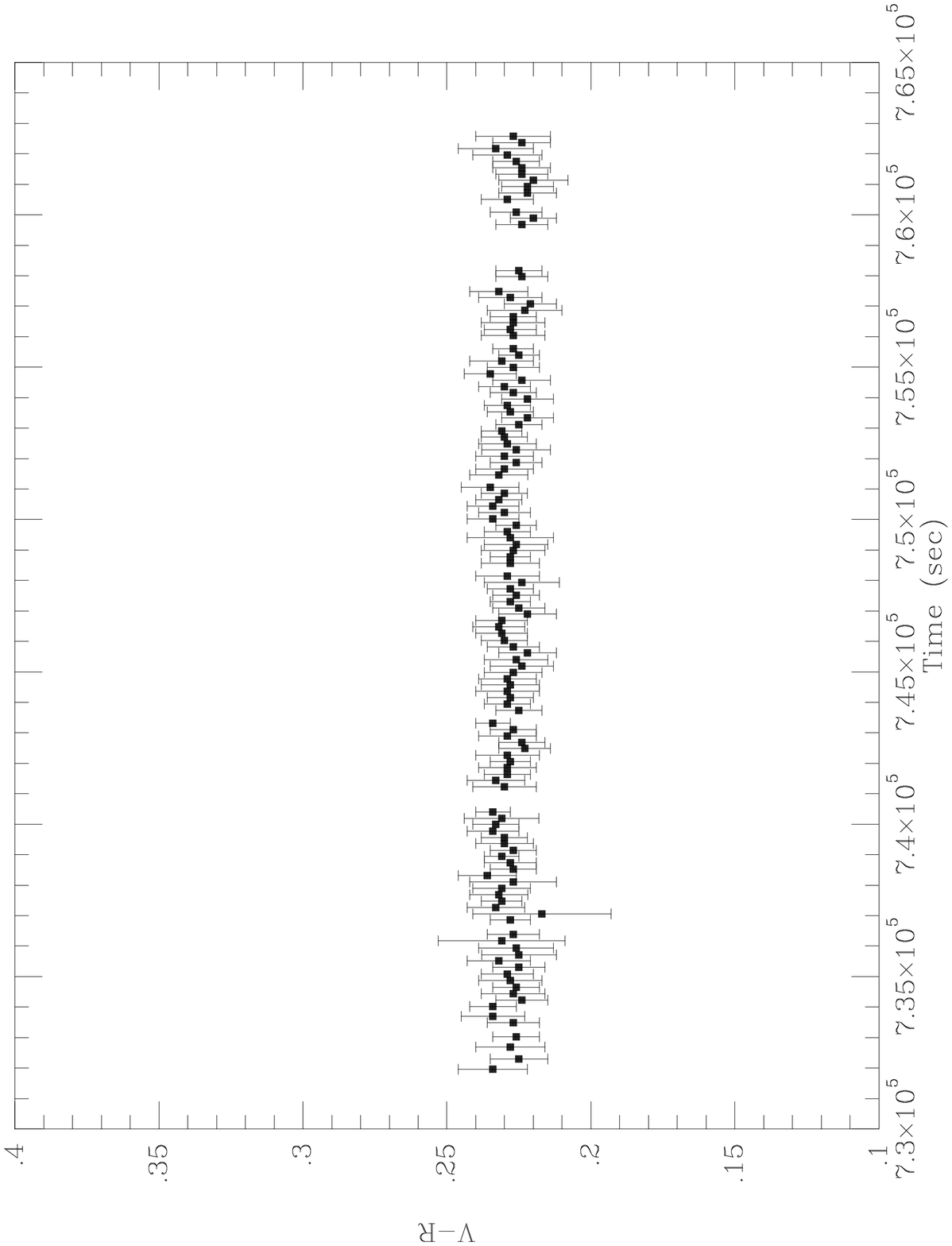}
\caption {The \vr\ color of PKS~2005--489 during the sixth night of the campaign. Time $T=0$ corresponds to
the beginning of the multiwavelength campaign on 19 August 2000 at 11:53:04 UTC.}
\label{fig-4}
\end{figure}

Figure~\ref{fig-2} shows the $V$ magnitude of PKS~2005--489 during the sixth night of the campaign, 27 August
2000.  Figures~\ref{fig-3} and~\ref{fig-4} show the \bv\ and \vr\ color variability respectively during this
time.  These figures are representative of the behavior of PKS~2005--489 during the entire campaign.  Despite
the precision of the photometry, there is no evidence for flux density or optical color variability
during any night of the campaign on timescales of hours or shorter.  Any variability on intraday timescales
is at a level well below the 1$\sigma$ RMS error of the individual measurements.  

Unfortunately the RXTE observations during the campaign are mostly insufficient to search for X-ray IDV.
Due to the failure of PCU0 on RXTE during the observations, as well as the faintness of PKS~2005--489 at X-ray
wavelengths during the campaign, the X-ray data had to be binned by 10~ksec to obtain statistically meaningful
flux density measurements \citep{rec02}.  However, the data are sufficient to show that the long-term variability behavior
in the X-ray is different than at optical wavelengths during the campaign.  As shown in Figure 1 of \citet{rec02}, the
X-ray flux doubles, with no spectral change, between 19--25 August, remains constant during 26--27 August and then
fades on 28 August.  If the stable X-ray period of 26--27 August corresponds to the stable optical period of 28
August this indicates a
$\sim$2 day lag between the optical and the X-ray.  If this is indeed the case, this lag is much longer than
observed in other blazars \citep{sam00,tan01,pia02,ulr97}.  Unfortunately the lack of detectable variability on short
timescales prevents a correlation of the optical and X-ray light curves and reduces this analysis to speculation.

\section{Discussion}

There is no evidence for IDV in PKS~2005--489 during our campaign, to a relative photometric precision of $\la 0.015$ magnitudes over a
6.2 day period.    Similarly, the optical monitoring of \citet{hei96} detected only modest variability in this source of
10.1\% over a 5.08 day period.  
Nor is there any evidence for optical color variability, although this is not unexpected as the synchrotron peak is at much higher energies.  Previous efforts have found little color variability in the optical and near-IR in PKS~2005--489 and other HBLs despite large variability in flux density \citep{tag01,fan99}.

There is no known evidence for optical IDV in this source at any epoch; which is not that unusual for HBLs, as they tend to show fewer instances of optical IDV, and with a shorter duty cycle, than RBLs \citep{hei98}.
However, intraday X-ray variability has been seen in PKS~2005--489 on one occasion, during the last two days of the 1998 September
campaign of \citet{sam00}, where oscillations of $\sim$20\% were seen on timescales of 6--8 hours.  Other campaigns have not seen X-ray IDV, e.g., \citet{per99b} and \citet{lam02}. 
Unfortunately PKS~2005--489 has not yet been monitored in the optical on short timescales during an X-ray
outburst; thus it is not known if optical IDV occurs in this object during a flare, or shortly thereafter. 

On longer timescales, a linear increase in optical flux density of 13\% was measured during the first 5.2 days of
this campaign, just a few days prior to the X-ray outburst on 2 September 2000.  This behavior has been seen in PKS~2005--489 before, when it was observed to brighten gradually by $\sim$0.5 magnitudes during the month prior to the X-ray flare in November 1998.

The behavior of PKS~2005--489 reported here and in \citet{rec02} is both similar to, and different
from, that seen in 1998, when massive flaring activity
was reported.  The flaring
activity in 1998 consisted of two peaks, in September \citep{sam00}
and November \citep{per99b}. Comparison of the X-ray lightcurves
during the two 1998 peaks shows that the 2--6~keV flux densities were within 20\% of
one another; however, very different behavior was seen above 6~keV, where the
flux density in November was several times that seen at the peak of the September flare. Moreover,
the character of the variability seen at those times was quite different,
with hard X-ray variability on 6--8 hour timescales recorded in September but
no intraday variability seen in October and November. Unfortunately not much
information exists on the optical lightcurve of PKS~2005--489 during those
times, but an inspection of data from the AAVSO (J. Maffei 2002, private
communication) as well as \citet{sca99} and other observers reveals
much gentler optical variability during late 1998, with a nearly constant
(variations $\la 0.2$ mag) optical magnitude in July-early October, and
then modest (0.7~magnitude) increases in late October and late December.  
Comparing these lightcurves to the models of \citet{chi99}
indicates that the two 1998 flares probably originated from the same physical
region, one in which both the particle injection timescale $t_{inj}$ and the 10--15~keV cooling timescale
 $t_{cool, \rm{10-15 keV}}$ were comparable
to the light crossing time $R/c$ of the varying region, because of (1) the small
shift seen in $\nu_{peak}$, (2) the symmetry of the lightcurve during the November flare, 
(3) the 4-day-long near-plateau in the 10--15~keV lightcurve, and
(4) the lack of response seen in the hard X-ray lightcurve until the November
flare, which probably re-accelerated to even higher energies particles that
had also been accelerated in September. 

By contrast, while the X-ray and optical lightcurves during our 2000 campaign
did show significant variability, for various reasons the data did not enable
us to model changes similar to those that occurred two years earlier. As
reported in \citet{rec02}, the X-ray lightcurve showed a smooth,
monotonic increase by nearly a factor 2 during 19-27~August, a decrease of
$\sim$40\% on 28~August, then roughly constant flux density between 29~August-1~September, 
and then a rapid increase in flux density 2-4~September, nearly
quadrupling the source's flux density within the final 2-day span.  Unfortunately,
however, the source was approximately a factor of 20 fainter than during
the 1998 flare.  Thus we were unable to observe significant X-ray spectral
changes during the 2000 campaign.  And then the sampling was too poor to ascertain the shape of the
start of the flare on 2 September.  

The optical lightcurve discussed here
shows a smooth, monotonic increase of about 13\% during 22-27~August.
However, as already discussed we have no optical data for the latter part of
the campaign when the largest X-ray excursions were seen.  The 1998 and 
2000 lightcurves differ in the magnitude of the X-ray varibility, which was over a factor 10 in hard X-rays during 1998, 
but was much less in 2000.  This by itself does not
yield significant constraints. But what is  interesting about
the 1998 and 2000 lightcurves of PKS~2005--489 is that in both cases, the
X-ray variability was of significantly greater amplitude and was more
violent, than that seen in the optical.  And no optical IDV was seen in
either year, with X-ray IDV being seen only in September 1998 but not during
the main flare two months later, although in 1998 the sampling was very poor
due to the poor visibility in September-December as well as scheduling constraints. This is in general
agreement with what is predicted for synchrotron models, which predict more
violent variability at higher frequencies, where the cooling time is shorter.  
However, without higher signal to noise we cannot use the 2000
X-ray lightcurves to place tighter constraints on the size of or physical
conditions within the flaring regions in the jet of PKS~2005--489.
Nor can we test Comptonization models with the available dataset because PKS~2005--489 was not
detected at TeV energies during the campaign; and the SED of PKS~2005--489 can be explained as synchrotron emission up to energies of 20 keV and higher \citep{tag01,pad01,rec02}.  

We also note that PKS~2005--489 was observed to be in an unusual state during this campaign, wherein its optical flux density was as
high as seen in any other epoch but its X-ray flux density was very faint compared to prior epochs.  For
comparison, the optical flux density for PKS~2005--489 during this campaign was 0.5--1.0 magnitudes brighter than
measured in the October-December 1998 campaign even though the X-ray flux density was in a low
state, with a 2--10~keV flux density comparable to that observed by \citet{sam94} and thirty times fainter than
the flare observed in the October-December 1998 campaign \citep{per99b}.  
%
%
%
%
%
In this campaign, and in the
October-December 1998 campaign, the observed gradual increase in optical flux density was accompanied by a similar
increase in X-ray flux density prior to an X-ray outburst, suggesting a rough correlation on the timescales of
weeks.  Due to a lack of small-scale structure in the X-ray and
optical lightcurves during the two campaigns, it is not possible to determine a more precise time offset between the two
wavebands.  However, the large discrepancy between the ratio of X-ray to optical flux density between the two epochs
indicates that the correlation is not long-term.  


Thus the relationship between the optical and X-ray emission in PKS~2005--489 is unclear.  As discussed above, the two 1998 X-ray flares are consistent  with simple particle injection models where $t_{inj}$ and $t_{cool, {\rm 10-15 keV}}$ are comparable to $R/c$ of the varying region.  As such, the optical peak due to synchrotron
cooling is expected to follow the X-ray peak.   In both 1998 and 2000 optical brightening occurs prior to the
X-ray flares; and unfortunately optical coverage after these flares is poor or unavailable.   On long timescales, the observed slow variability can be consistent with particle injection models if
the region in question is large and/or the magnetic field is weak.  However, the large discrepancy between the
ratio of X-ray to optical flux density observed in this epoch and that observed in October 1998 indicates a lack of
long-term correlation between the two wavebands; and it illustrates that simple particle injection events are likely an insufficient explanation for the jet emission mechanism.

\section{Conclusions}

We have presented rapid, multi-color optical monitoring carried out for a multiwavelength campaign during 22-27 August 2000 for the high-energy peaked BL Lac object PKS~2005--489.  These observations are part of the
multifrequency monitoring campaign reported in \citet{rec02}.   During this campaign, PKS~2005--489 was observed to be in an unusual state, bright in the optical but faint in the X-ray, just prior to an X-ray flare that began on 2 September 2000.  A linear increase in optical flux density was observed during the first 5.2 days of the campaign.  However, no optical flux density or color IDV was detected, although PKS~2005--489 was  observed to have a flatter optical spectrum than in prior epochs.  PKS~2005--489 was too faint during this campaign to search for IDV at X-ray energies.

The two 1998 X-ray flares of PKS~2005--489 probably originated from the same physical region, one in which both the particle injection timescale and the 10--15~keV cooling timescale were comparable to the light crossing time of the varying region.  The coverage of our 2000 campaign prevents the study of the flare on 2 September; however, we note that, in both the 1998 and 2000, the X-ray variability was of significantly greater amplitude, and was more violent, than that seen in the optical.  And no optical IDV was seen in either year.  This is in general agreement with what is predicted for synchrotron models, which predict more violent variability at higher frequencies, where the cooling time is shorter.

The large discrepancy between the ratio of X-ray to optical flux density observed during this campaign and that observed in prior epochs indicates a lack of long-term correlation between the two wavebands; and it illustrates that simple particle injection models for the jet emission mechanism are insufficient to explain the long-term variability behavior of this object.
The behavior of PKS~2005--489 during this and prior epochs shows a variety of violent and complex behavior, encouraging future campaigns.  Twenty-three in-field standard stars are identified for future optical monitoring programs.

\acknowledgments

E. Perlman acknowledges interesting coversations with M. Chiaberge about 
the interpretation of multi-band blazar variability.  We also wish to thank the referee for insightful comments which
have improved the quality of this paper.

\clearpage


%
%


\clearpage

\clearpage

\clearpage

\clearpage





\clearpage

\begin{deluxetable}{cccccc}
\tablecaption{In-Field Standard Stars \label{tbl-1}}
\tablewidth{0pt}
\tablehead{\colhead{Star} & \colhead{RA(J2000)} & \colhead{Dec} & \colhead{$V$} & \colhead{\bv} &
\colhead{\vr}}

\startdata

A	&	20:09:23.4 &	--48:44:59 & 12.89	&	0.60	&	0.34	\\
B	&	20:09:11.2 &	--48:45:16 & 14.96	&	1.16	&	0.70	\\
C	&	20:09:12.0 &	--48:45:01 & 14.97	&	0.95	&	0.57	\\
D	&	20:08:58.7 &	--48:45:18 & 15.45	&	0.73	&	0.38	\\
E	&	20:09:02.3 &	--48:45:35 & 13.11	&	0.48	&	0.26	\\
F	&	20:09:22.9 &	--48:45:39 & 14.18	&	0.78	&	0.44	\\
G	&	20:09:52.0 &	--48:46:17 & 14.99	&	0.70	&	0.38	\\
H	&	20:09:38.6 &	--48:46:18 & 13.59	&	0.93	&	0.52	\\
I	&	20:09:59.9 &	--48:46:31 & 14.97	&	0.71	&	0.39	\\
J	&	20:09:04.2 &	--48:46:33 & 15.13	&	0.69	&	0.40	\\
K	&	20:10:02.6 &	--48:46:32 & 13.88	&	0.50	&	0.28	\\
L	&	20:09:19.0 &	--48:46:44 & 14.41	&	0.97	&	0.52	\\
M	&	20:09:00.5 &	--48:48:23 & 14.81	&	0.82	&	0.42	\\
N	&	20:09:07.1 &	--48:48:27 & 15.09	&	0.65	&	0.36	\\
O	&	20:08:59.1 &	--48:48:34 & 14.01	&	0.74	&	0.40	\\
P	&	20:09:53.0 &	--48:48:42 & 14.76	&	0.85	&	0.48	\\
Q	&	20:09:48.7 &	--48:48:52 & 12.14	&	0.55	&	0.30	\\
R	&	20:09:01.4 &	--48:48:59 & 15.04	&	1.01	&	0.58	\\
S	&	20:09:00.6 &	--48:49:06 & 12.68	&	1.19	&	0.63	\\
T	&	20:09:40.0 &	--48:50:22 & 14.92	&	0.97	&	0.53	\\
U	&	20:09:17.7 &	--48:50:57 & 14.88	&	0.57	&	0.33	\\
V	&	20:09:59.3 &	--48:51:23 & 15.41	&	0.73	&	0.41	\\
W	&	20:09:50.1 &	--48:51:33 & 14.67	&	0.60	&	0.34	\\

\enddata
\end{deluxetable}

\clearpage


\end{document}